\documentstyle[preprint,aps,epsf,floats,psfig]{revtex}

\tighten

\newcommand{\beq}{\begin{equation}}
\newcommand{\eeq}{\end{equation}}
\newcommand{\beqa}{\begin{eqnarray}}
\newcommand{\eeqa}{\end{eqnarray}}

\newcommand{\ga}{g_A}
\newcommand{\gas}{g_A^2}

\newcommand{\mpi}{m_\pi}
\newcommand{\mpis}{m_\pi^2}

\newcommand{\lsim}{\raisebox{-0.7ex}{$\stackrel{\textstyle <}{\sim}$ }}

\def\sss{{\scriptscriptstyle}}

\def\si{{}^1\kern-.14em S_0}
\def\siii{{}^3\kern-.14em S_1}
\def\piii{{}^3\kern-.14em P_1}
\def\diii{{}^3\kern-.14em D_1}

\begin{document}

\begin{titlepage}

\hfill{NT@UW-02-015}

\vspace{0.5cm}

\begin{center}
{\Large\bf Variation of Fundamental Couplings\\ \vspace{0.2cm}
and Nuclear Forces}

\vspace{1.2cm}

{\bf Silas R.~Beane and Martin J.~Savage}

\vspace{0.2cm}
{\it
Department of Physics,
University of Washington,\\
Seattle, WA 98195-1560}
\end{center}

\vspace{0.1cm}

\begin{abstract}
  
  The dependence of the nuclear force on standard model parameters plays an
  important role in bounding time and space variations of fundamental couplings
  over cosmological time scales. We discuss the quark-mass dependence of
  deuteron and di-neutron binding in a systematic chiral expansion. The leading
  quark-mass dependence of the nuclear force arises from one-pion exchange and
  from local quark-mass dependent four-nucleon operators with coefficients that
  are presently unknown. By varying these coefficients while leaving nuclear
  observables at the physical values of the quark masses invariant, we find
  scenarios where two-nucleon physics depends both weakly and strongly on the
  quark masses.  While the determination of these coefficients is an exciting
  future opportunity for lattice QCD, we conclude that, at present, bounds on
  time and space variations of fundamental parameters from the two-nucleon
  sector are much weaker than previously claimed. This brings into question the
  reliability of coupling-constant bounds derived from more complex nuclei and
  nuclear processes.

\end{abstract}

\vspace{2cm}
\vfill
\end{titlepage}
\setcounter{page}{1}


\section{Introduction}

The recent observation suggesting that the fine-structure constant was smaller
in the past~\cite{dalpha} than it is today has led to renewed interest in the
idea of using time (and space) variation of fundamental parameters as a probe
of high-energy physics.  Based on the principle that all that is not
forbidden is mandatory, time variation in the electromagnetic force is perhaps
not surprising. Moreover, one would expect that the other forces of nature
vary with time as well.  This idea is quantifiable by assuming a suitable grand
unified theory (GUT) at a scale $M_{\sss GUT}$ or a particular brane-world
scenario~\cite{CF,matt,Chacko:2002mf}. 
The assumption that the short-distance couplings of the theory are
related at all times leads to relations between the variation of the strength
of the electromagnetic interaction, $\alpha_{\rm em}$, and the variation of
other standard model parameters, such as the light quark masses, $m_q$.  As a
change in $m_q$ and $\alpha_{\rm em}$ will naively lead to a change in the
positions of nuclear energy levels, special interest has been paid to the 
light-element abundances predicted by big bang nucleosynthesis (BBN) and also to the
abundance of isotopes produced by the Oklo ``natural reactor'' in the hope that
these abundances can be used to constrain high-energy
physics~\cite{CF,matt,KPW,Barrow,Dixit,campbell,Damour,agrawal,hummer,Fujii,Shera,BIR,Csoto,Chiba,dent,fair,Dolgov,IK,shuryak,olive,uzan}.

In this work we will critically analyze the two-nucleon sector using an
effective field theory (EFT) that respects the approximate 
$SU(2)_L\otimes SU(2)_R$ chiral symmetry of
QCD and has consistent power-counting~\cite{We90,KSWa,KSWb,Be01}.  This theory
allows for a systematic study, consistent with QCD, of the bound-state in the
$\siii-\diii$ coupled-channels, the deuteron, and the scattering amplitude in
the $\si$ channel. 
This is the first study of the $m_q$-dependence of these quantities that
is {\it complete} at next-to-leading order (NLO) in a consistent EFT.
Unfortunately, due to the current lack of understanding of the
low-energy behavior of QCD, only weak bounds, if any, can be placed
on the  time-dependence of fundamental couplings from the two-nucleon sector.
We will explicitly demonstrate this by finding plausible values for couplings in the
EFT for which the deuteron binding energy varies relatively little, and for
which the di-neutron system in the $\si$ channel is unbound, over a relatively 
wide range of quark masses.

The deuteron is an interesting object to study for a variety of reasons.
From a fundamental point of view it is quite intriguing as it is bound by only 
$B_{\rm d} = 2.224644\pm 0.000034~{\rm  MeV}$, which is much smaller than the typical scale
of strong interactions, $\Lambda_{\sss QCD}$.
Clearly, in order to arrive at such a small binding energy, fine-tunings are
involved~\footnote{In the pionless EFT one finds that both the $\si$ and the
$\siii -\diii$ coupled-channels are very close to an unstable IR fixed point of
the renormalization group. At this fixed point, the scattering lengths are 
infinite~\cite{KSWb,Birse}.} and consequently naive dimensional analysis (NDA) as applied to the
single-nucleon sector or the mesonic sector, which we will call naive naive
dimensional analysis (N$^2$DA), on this system is doomed
to fail.  Unfortunately, all previous attempts to extract bounds on the
variations of fundamental couplings from the deuteron, and in general the
two-nucleon sector, have implemented N$^2$DA.
From a more phenomenological standpoint, the smallness of $B_{\rm d}$ plays
a key role in the synthesis of light elements in BBN.
The impressive agreement between the predictions of BBN and observation
suggests that new physics that would have significantly modified the deuteron 
at the time of BBN is absent.
The $\si$ channel is quite similar to the $\siii-\diii$ coupled-channels in
one important way, its scattering length is unnaturally large,
$a^{(\si)}=-23.714\pm 0.013~{\rm fm}$.
While there is no bound state in this channel for the physical values of the
light-quark masses, a small increase in the strength of
the nuclear force would bind two nucleons in this channel.
The existence of a bound state in this channel, e.g. a di-neutron, $nn$,
in the nucleosynthesis epoch would be quite profound and could substantially 
modify the predictions of BBN.

\section{EFT for Two Nucleons}

During the last decade there has been a significant effort to construct an EFT
to describe nuclear physics.  While it is straightforward to write down all
possible terms in the effective Lagrange density for two or more nucleons,
arriving at the correct power-counting proved to be a difficult task.
Weinberg's (W) original proposal~\cite{We90} for an EFT describing
multi-nucleon systems was to determine the nucleon-nucleon (NN) potentials
using the organizational principles of the well-established EFT's describing
the meson-sector and single-nucleon sector (chiral perturbation theory,
$\chi$PT), and then to insert these potentials into the Schr\"odinger equation
to solve for NN wavefunctions.  Observables are computed as matrix elements of
operators between these wavefunctions.  W power-counting has been extensively
and successfully developed during the past decade to study processes in the
few-nucleon systems.  This method is
intrinsically numerical and is similar in spirit to traditional nuclear-physics
potential theory.  Unfortunately, there are formal inconsistencies in W
power-counting~\cite{KSWa}, in particular, divergences that arise at leading
order (LO) in the chiral expansion cannot be absorbed by the LO operators.
Problems persists at all orders in the chiral expansion, and the correspondence
between divergences and counterterms appears to be lost, leading to
uncontrolled errors in the predictions for observables.  This formal issue was
partially resolved by Kaplan, Savage and Wise (KSW) who introduced a power-counting in
which pions are treated perturbatively~\footnote{ In KSW power-counting the momentum-independent
  four-nucleon operators are LO and are resummed to yield the LO
  scattering amplitude, while pions are NLO and treated in perturbation
  theory.}~\cite{KSWb}.  The NN phase-shifts and
mixing angle in the $\si$ and $\siii-\diii$ coupled-channels have been computed
to next-to-next-to-leading order (N$^2$LO) in the KSW expansion by Fleming,
Mehen and Stewart (FMS)~\cite{FMS} from which it can be concluded that the KSW
expansion converges slowly in the $\si$ channel and does not converge in the
$\siii-\diii$ coupled-channels. Therefore, neither W or KSW power-counting 
provide a complete description of nuclear interactions 
(for recent reviews see Ref~\cite{Be00,pablito,danielito,ulfito,parkito}).

The problems with W and KSW power-counting appear to have been resolved in the
work by Bedaque, van Kolck and the authors~\cite{Be01}, which from this point
on we will refer to as BBSvK.  It was realized in FMS that the contributions to
the amplitude that lead to non-convergence in the $\siii-\diii$
coupled-channels 
persist in the chiral limit (it is the chiral limit of iterated
one-pion-exchange (OPE) that is troublesome).  Therefore, in BBSvK
power-counting 
the scattering amplitude is an expansion about the chiral limit.
This recovers KSW power-counting in the $\si$ channel, where FMS found it to be
slowly converging.  However, in the $\siii-\diii$ coupled-channels, the chiral limit
has contributions from both local four-nucleon operators and from the chiral
limit of OPE.  It is these two contributions that must be
resummed using the Schr\"odinger equation to provide the LO scattering
amplitude in the $\siii-\diii$ coupled-channels~\footnote{ A technical issue
  that remains to be solved is that while dimensional regularization can be
  used to regulate the $\si$ channel, so far no one has managed to
  dimensionally regulate the $\siii-\diii$ coupled-channels due to the presence
  of the $1/r^3$ component of the nucleon-nucleon potential.  Therefore, in this
  work we will dimensionally regulate the $\si$ channel and use position-space
  square-well regularization in the $\siii-\diii$ coupled-channels.  It has
  been shown that a singular $1/r^n$ potential ($n\geq 2$), can be regulated and renormalized by a
  single momentum-independent square-well~\cite{kiddies}.  }.

The LO Lagrange density describing the 
single-nucleon sector and the pseudo-Goldstone bosons in two-flavor QCD is
\begin{eqnarray}
{\cal L} & = & 
{1\over 8}{(f_\pi^{(0)})^2}\, {\rm Tr}\left[\ 
\partial^\mu\Sigma^\dagger\partial_\mu\Sigma\ \right]
\ +\ \lambda {\rm Tr}\left[\ m_q\Sigma^\dagger + m_q\Sigma\ \right]
\nonumber\\ & &\ +\  N^\dagger \left(\ i\partial_0 + {\nabla^2/ 2 M_N^{(0)}} \ \right) N
\ +\ g_A^{(0)} N^\dagger \sigma\cdot {\bf A} N 
\ \ \ ,
\label{eq:onebody}
\end{eqnarray}
where $f_\pi^{(0)}$ is the pion decay constant in the chiral limit,
$M_N^{(0)}$ is the nucleon mass in the chiral limit,
$g_A^{(0)}$ is the axial coupling constant in the chiral limit,
$\Sigma$ is the exponential of the pion field, 
and ${\bf A}$ is the axial-vector field.
The properties of the nucleons and mesons have been studied 
in the first few orders
of the chiral expansion (a good review can be found in
Ref.~\cite{ulfioffe}).
If one is interested in the $m_q$-dependence of the nuclear force, as we are,
one needs to have the chiral expansion for the nucleon mass, for the 
axial coupling and for the pion decay constant up to NLO.  
Each of these observables has been studied extensively, the results of which
can be found in Refs.~\cite{ulfioffe,GaLe84,CoGaLe01,FeMe00},
\begin{eqnarray}
f_\pi & = & f_\pi^{(0)} \left[ 1 + { m_\pi^2\over 8\pi^2 (f_\pi^{(0)})^2} 
\overline{l}_4\ +\ {\cal O}\left(m_\pi^4\right)\ \right]
\nonumber\\
M_N & = & M_N^{(0)} - 4 m_\pi^2 c_1 + {\cal O}\left(m_\pi^3\right)
\nonumber\\
g_A & = & g_A^{(0)} \left[ 1 - 
{ 2 (g_A^{(0)})^2+1
\over 8\pi^2 (f_\pi^{(0)})^2} 
m_\pi^2\log\left({m_\pi^2\over{\lambda^2}}\right)+{\cal O}(m_\pi^2)\right]
\ \ \ ,
\label{eq:SNparams}
\end{eqnarray}
where $\overline{l}_4=4.4\pm 0.2$~\cite{GaLe84,CoGaLe01}, 
$c_1 \sim -1~{\rm  GeV}^{-1}$~\cite{ulfioffe} are $m_q$-independent constants, 
and $m_\pi=139~{\rm MeV}$ has been used to determine the constants in the 
chiral limit. We use $g_A=1.25$, $M_N=(M_n+M_p)/2$ and $f_\pi =135~{\rm MeV}$.
We have retained only the leading chiral-logarithmic contribution to $g_A$, and have
chosen a renormalization scale of $\lambda=500~{\rm MeV}$. This point requires
discussion. Extraction of the counterterm relevant to the $m_q$-dependence of $\ga$ at
one-loop order presently yields $\lambda\sim 100~{\rm MeV}$ (see Ref.~\cite{ulfioffe} 
and references therein). As this anomalously small value seemingly indicates a
breakdown of the chiral expansion, we assume that this is a problem with the extraction
and use a natural value of $\lambda = 500~{\rm MeV}$ for this analysis. This, of course,
introduces an uncertainty at NLO in the EFT calculation.

The interactions between two nucleons arise from pion exchange, resulting from
the Lagrange density in eq.~(\ref{eq:onebody}), and also from local 
four-nucleon interactions, which for s-wave interactions result from a Lagrange
density of the form,
\begin{eqnarray}
{\cal L} & = & 
-{1\over 2} C_S \left( N^\dagger N\right)^2
\ -\ {1\over 2} C_T \left( N^\dagger \sigma^i N\right)^2
\nonumber\\ & & 
-{1\over 2} D_{S1} \left( N^\dagger {\cal M}_{q+} N\right)
\left( N^\dagger N\right)
-{1\over 2} D_{T1} \left( N^\dagger {\cal M}_{q+} \sigma^i N\right)
\left( N^\dagger \sigma^i N\right)
\nonumber\\ & & 
-{1\over 2} D_{S2} \left( N^\dagger N\right)
\left( N^\dagger N\right) {\rm Tr}\left[ {\cal M}_{q+} \right]
-{1\over 2} D_{T2} \left( N^\dagger \sigma^i N\right)
\left( N^\dagger \sigma^i N\right){\rm Tr}\left[ {\cal M}_{q+} \right]
\ +\ ...
\ \ \ ,
\label{eq:twobody}
\end{eqnarray}
where
${\cal M}_{q+} = {1\over 2}\left(\ \xi^\dagger m_q \xi^\dagger + \xi m_q
  \xi \ \right)$, and $m_q={\rm diag}(m_u,m_d)$. 
The ellipses denote operators
involving two derivatives~\cite{carlos}, which are the same order in the power-counting as
a single insertion of $m_q$, and also higher-dimension operators.
The $m_q$-independent coefficients $C_i$, $D_i$ and so forth, are to be determined from
experimental data.  An important point to notice is that 
one cannot separate the contributions from the $C_i$ and the $D_i$ using NN scattering data alone,
as both operators are momentum independent.
However, the $D_i$ operators contribute to interactions between 
two nucleons and two or more pions, while the $C_i$ operators only contribute
to interactions between two nucleons.

\section{The $\si$ Channel and the Di-neutron}

One might imagine that grave violence would be done to the predictions of 
BBN if a di-neutron or di-proton were to be stable during the nucleosynthesis
epoch.
Thus it is interesting to know just how much the couplings in the 
Lagrange density describing the interactions between two nucleons 
in the $\si$ channel can change before a di-neutron state becomes 
stable~\footnote{In this initial study, we do not consider the di-proton system due to the 
coulomb potential that further destabilizes it. However, for a 
meaningful study of constraints from BBN in this channel, the 
possibility of $nn$, $np$ and $pp$ bound states would have to 
be considered.}.
However, from the point of view of setting bounds on the variation of
fundamental couplings, what is important to know is if there is a set of
``reasonable'' couplings for which the scattering length remains unnaturally large and  
the di-neutron is unbound over a large range of pion masses.
In this situation the pion mass can vary substantially, yet
still not significantly modify the physics of two nucleons in the $\si$
channel.  Indeed, such a parameter set exists.

\begin{figure}[!ht]
\centerline{{\epsfxsize=2.5in \epsfbox{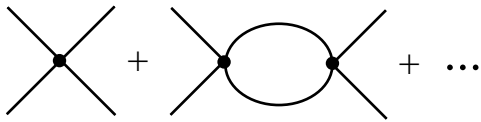}} }
\vskip 0.15in
\noindent
\caption{LO contribution to the scattering amplitude in the $\si$
  channel. The small solid circle denotes an insertion of $C_0$.}
\label{fig:1S0LO}
\vskip .2in
\end{figure}
\begin{figure}[!ht]
\centerline{{\epsfxsize=4.5in \epsfbox{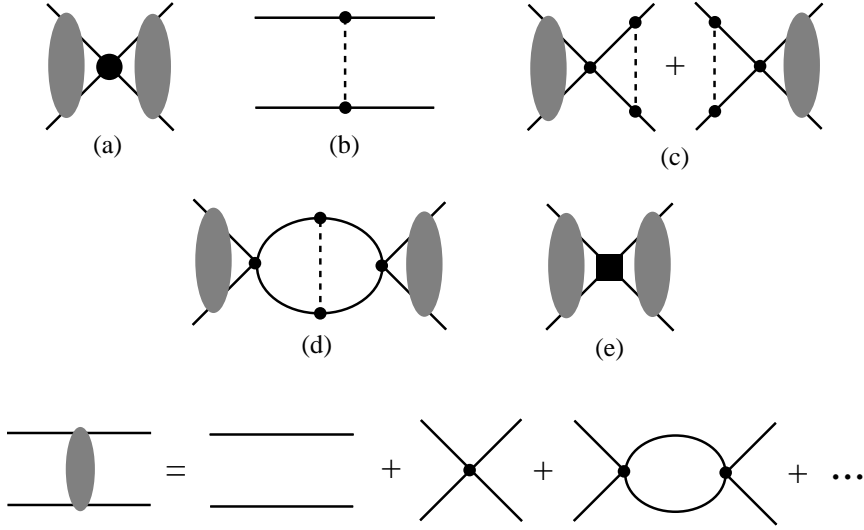}} }
\vskip 0.15in
\noindent
\caption{NLO contributions to the scattering amplitude in the
  $\si$ channel. The small solid circles denote an insertion of $C_0$ or $g_A$.
    Dashed lines are pions and the large solid circle (square)
  corresponds to an insertion of $C_2$ ($D_2$).}
\label{fig:1S0NLO}
\vskip .2in
\end{figure}

The scattering amplitude for two nucleons in the $\si$ channel has been
computed out to N$^2$LO~\cite{KSWb,FMS},
but for our purposes it will be sufficient to work with the amplitude at
NLO~\cite{KSWb}.
The scattering amplitude thus has an expansion of the form
\begin{equation}
{\cal A} = {\cal A}_{-1} +  {\cal A}_0 + {\cal A}_1 +\ldots
\end{equation}
where ${\cal A}_n$ is of order $Q^n$ in the dual $m_q$ and momentum expansions.
The LO amplitude arises from the diagrams in Fig.~\ref{fig:1S0LO} and is given by 

\begin{eqnarray}
{\cal A}_{-1} & = & {- C_0\over \left[1 + {C_0 M\over 4\pi} (\mu + ip)\right]}
\ \ \ ,
\label{eq:ampLO}
\end{eqnarray}
while the NLO amplitude arises from the diagrams in Fig.~\ref{fig:1S0NLO} and is given by the sum of 
\begin{eqnarray}
{\cal A}_0^{(a)} & = &  { - C_2 p^2\over \left[1 + {C_0 M\over 4\pi}
(\mu + ip)\right]^2}
\nonumber\\
 {\cal A}_0^{(b)} &=&  \left({g_A^2\over 2f_\pi^2}\right) \left(-1 + {m_\pi^2\over
4p^2} \ln \left( 1 + {4p^2\over m_\pi^2}\right)\right)
\nonumber\\
{\cal A}_{0}^{(c)} &=& {g_A^2\over f_\pi^2} \left( {m_\pi M{\cal A}_{-1}\over 4\pi}
\right) \Bigg( - {(\mu + ip)\over m_\pi}
+ {m_\pi\over 2p} \left[\tan^{-1} \left({2p\over m_\pi}\right) + {i\over 2} \ln
\left(1+ {4p^2\over m_\pi^2} \right)\right]\Bigg)
\nonumber\\
{\cal A}_0^{(d)} &=& {g_A^2\over 2f_\pi^2} 
\left({m_\pi M{\cal A}_{-1}\over 4\pi}\right)^2 
\Bigg(1 -\left({\mu + ip\over m_\pi}\right)^2
+ i\tan^{-1} \left({2p\over m_\pi}\right) - {1\over 2} \ln
\left({m_\pi^2 + 4p^2\over\mu^2}\right) \Bigg)
\nonumber\\
{\cal A}_0^{(e)} &=&  {-D_2 m_\pi^2\over\left[1 + {C_0 M\over 4\pi} (\mu +
ip)\right]^2}
\ \ \ ,
\label{ampNLO}
\end{eqnarray}
where we have chosen to work in the isospin limit, $m_u=m_d$, and
we have turned off the electromagnetic interaction.
These amplitudes are manifestly renormalization-scale independent
order-by-order in the EFT expansion.
The coefficients 
$C_0$, $D_2$ and $C_2$ are the combinations of couplings from
the Lagrange density in eq.~(\ref{eq:twobody}) appropriate for the $\si$
channel.  
$C_0$ corresponds to a four-nucleon operator that is independent of $m_q$ and
momentum,
$D_2$ corresponds to a four-nucleon operator that has a 
single insertion of $m_q$ and no derivatives, and 
$C_2$ corresponds to a four-nucleon operator that is independent of $m_q$
and has two  derivatives.
The quantity $\mu$ is the renormalization scale, and we have used 
dimensional regularization and  the
power-divergence subtraction procedure (PDS)~\cite{KSWb} to renormalize the
theory; $p$ is the magnitude of the three 
momentum of each nucleon in the center-of-mass frame.
In addition, we have used the LO relation between the quark masses and the 
square of the pion mass so that the amplitude at NLO is written entirely in
terms of $m_\pi$.
It is important to note that the operator with coefficient $D_2$ is required
at NLO.  It is this operator that absorbs the scale
dependence of ${\cal A}_0^{(d)}$.

\begin{figure}[htb]
\vspace{-0.2cm}
\centerline{\psrotatefirst
\psfig{file=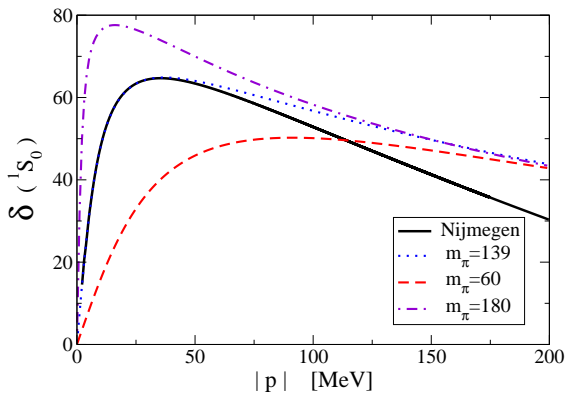,width=3.2in,angle=-90}}
\vspace{0.3cm}
\centerline{\parbox{14cm}{
\caption[fig1]{\label{fig:singphase}
The phase-shift, $\delta^{(\si)}$, as a function of momentum, $|{\bf p}|$,
for pion masses of $m_\pi=60~{\rm MeV}$ (dashed), 
$m_\pi^{\rm PHYS}$ (dotted) and $180~{\rm MeV}$ (dot-dashed),
for the couplings in eq.~(\protect\ref{eq:singpars}).
The solid curve corresponds to the results of the 
Nijmegen partial-wave analysis~\protect\cite{Nijmegen}.
}}}
\vspace{0.2cm}
\end{figure}

From the NLO amplitude it is easy to construct $p\cot\delta^{(\si)}$,
which has  a well-behaved power-series expansion for $p < m_\pi/2$, and thus we
can determine a linear combination of $C_0$ and $D_2$ in terms of the
scattering length $a^{(\si)}$, and $C_2$ in terms of the effective range,
$r^{(\si)}=2.73\pm 0.03~{\rm fm}$, at the physical value of the pion mass.
Once these parameters are fixed, the scattering length, effective range and
phase-shift can be determined  as a function of the pion mass. The scattering
length is given by

\begin{eqnarray}
\frac{1}{a^{(\si)}}=\gamma +\frac{g_A^2 M_N}{8\pi f_\pi^2}\left[ m_\pi^2\,
  \log{\left({\mu\over m_\pi}\right)}+ (\gamma - m_\pi )^2-(\gamma - \mu )^2 \right] 
-\frac{M_N m_\pi^2}{4\pi}(\gamma -\mu )^2 \ D_2  \ \ ,
\label{eq:scattlength}
\end{eqnarray}
where $\gamma =\mu +4\pi /M C_0$.
\begin{figure}[htb]
\vspace{-0.2cm}
\centerline{\psrotatefirst
\psfig{file=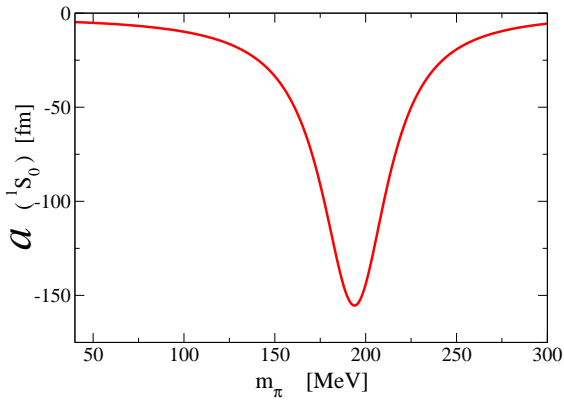,width=3.2in,angle=-90}}
\vspace{0.3cm}
\centerline{\parbox{14cm}{
\caption[fig1]{\label{fig:singscatt}
The scattering length in the $\si$ channel (in ${\rm fm}$'s)
as a function of the pion mass,
for the couplings given in eq.~(\protect\ref{eq:singpars}).
As the scattering length is negative over this entire region, the di-neutron
is unbound.
}}}
\vspace{0.2cm}
\end{figure}
Given that it is only a combination of $C_0$ and $D_2$ that can be fixed 
from NN scattering, and at present $D_2$ cannot be isolated with pionic
processes, unique values for $D_2$ or $C_0$ that are determined
experimentally do not exist.  
Thus, one must explore a ``reasonable'' range of values for $D_2$ and $C_0$ 
in order to gain some understanding of how the $\si$ channel is modified
away from the physical values of the quark masses.
It is quite straightforward to identify a set of couplings,

\begin{eqnarray}
C_0 (m_\pi^{\rm PHYS}) & = & -4.09~{\rm fm}^2
\quad , \quad 
D_2 (m_\pi^{\rm PHYS}) \ = \ 0.50~{\rm fm}^4
\quad , \quad 
C_2 (m_\pi^{\rm PHYS}) \ =\  4.94~{\rm fm}^4
\ \ \ ,
\label{eq:singpars}
\end{eqnarray}
for which the scattering length in the $\si$ channel remains 
unnaturally large and negative over a wide range of pion masses, as shown in 
Fig.~\ref{fig:singscatt}, and hence a di-neutron 
is not stable in this region~\footnote{For convenience, we plot observables as a function of
  $m_\pi$ rather than $m_\pi^2\propto m_q$, which would be more natural from the standpoint
  of QCD.}.
We have extracted the $C_i$ and $D_2$ using eq.~(\ref{eq:scattlength}) and an analogous
equation for the effective range, with $g_A$, $M_N$ and $f_\pi$ set to their chiral limit
values in eq.~(\ref{eq:SNparams}).
In order to determine the impact of a varying pion mass on nuclei more complex
than the deuteron, it is useful to know the behavior of the phase-shift as a
function of the pion mass over a relatively wide range of momentum.
In Fig.~\ref{fig:singphase} we show the $\si$ phase-shift 
for $m_\pi=60~{\rm MeV}$, $m_\pi^{\rm PHYS}$ and $180~{\rm MeV}$,
for the couplings in eq.~(\ref{eq:singpars}).
One notices that the NLO phase-shift does not reproduce the results of the 
Nijmegen partial-wave analysis~\protect\cite{Nijmegen} above 
$\sim 50~{\rm MeV}$, even at the physical
value of the pion mass~\footnote{For a detailed discussion of the $\si$ channel
description with EFT, see Refs.~\cite{KSWb,FMS,cohen}.}. This is because we have fixed the parameters,
$C_0$, $D_2$ and $C_2$ to the scattering length and effective range.
However, this means that the dual chiral and momentum
expansion of the EFT at NLO has been matched
to the momentum expansion of $p\cot\delta^{(\si)}$.
Consequently, both the scattering length and the effective range have their own
chiral expansions that we are truncating at NLO.
Thus we do not expect to reproduce the phase-shift exactly, but should be
perturbatively close, as can be seen to be the case in 
Fig.~\ref{fig:singphase}.
Rather than compare the phase-shifts with the 
Nijmegen partial-wave analysis~\protect\cite{Nijmegen},
it is perhaps more informative to compare the phase-shifts for 
$m_\pi=60~{\rm MeV}$ and $180~{\rm MeV}$ with those at 
$m_\pi=m_\pi^{\rm PHYS}$. 
Thus we have identified a set of NLO couplings for which the $\si$ 
channel is quite insensitive to moderate changes in the pion mass.

\begin{figure}[htb]
\centerline{\psrotatefirst
\psfig{file=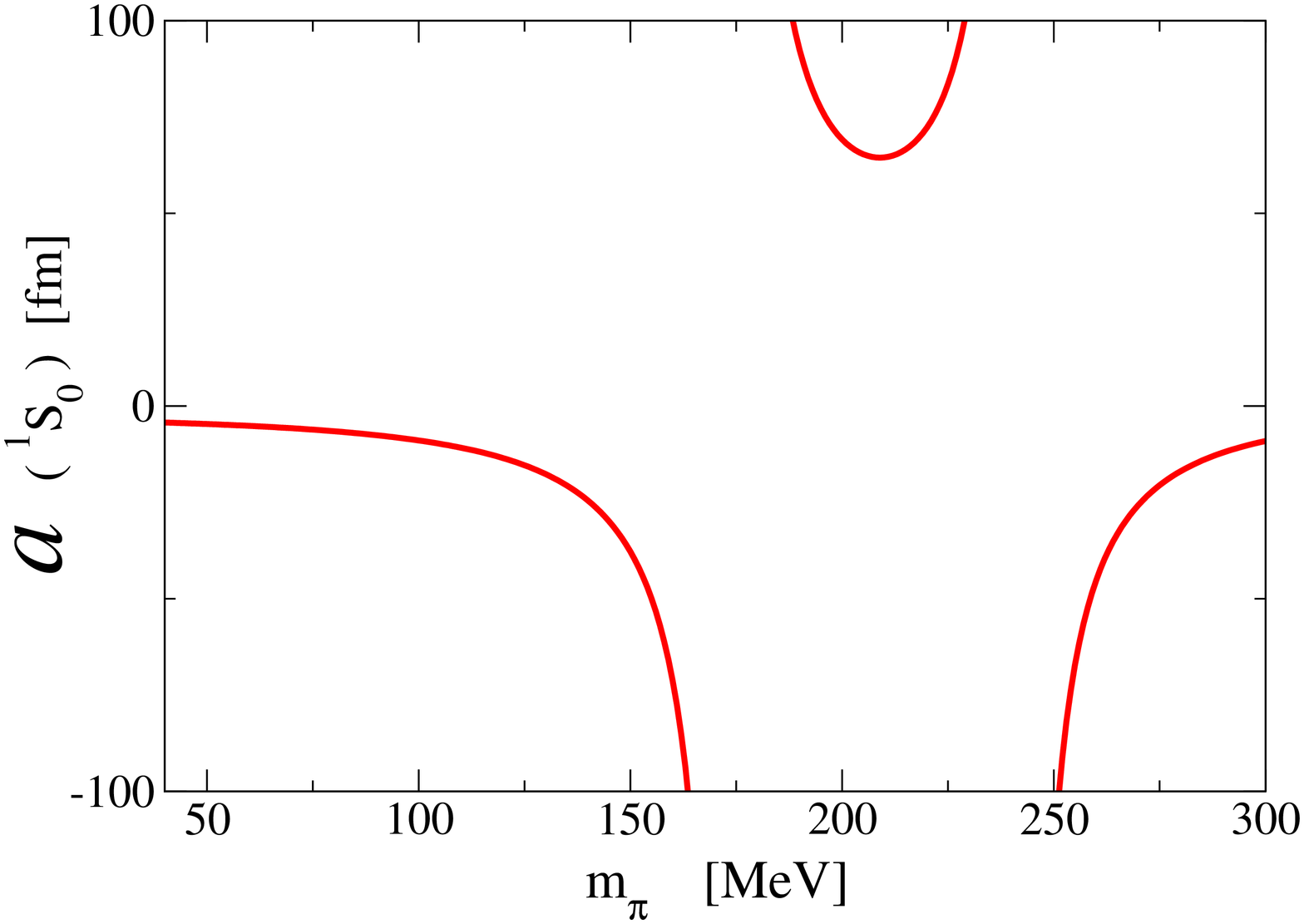,width=3.2in,angle=-90}
\psfig{file=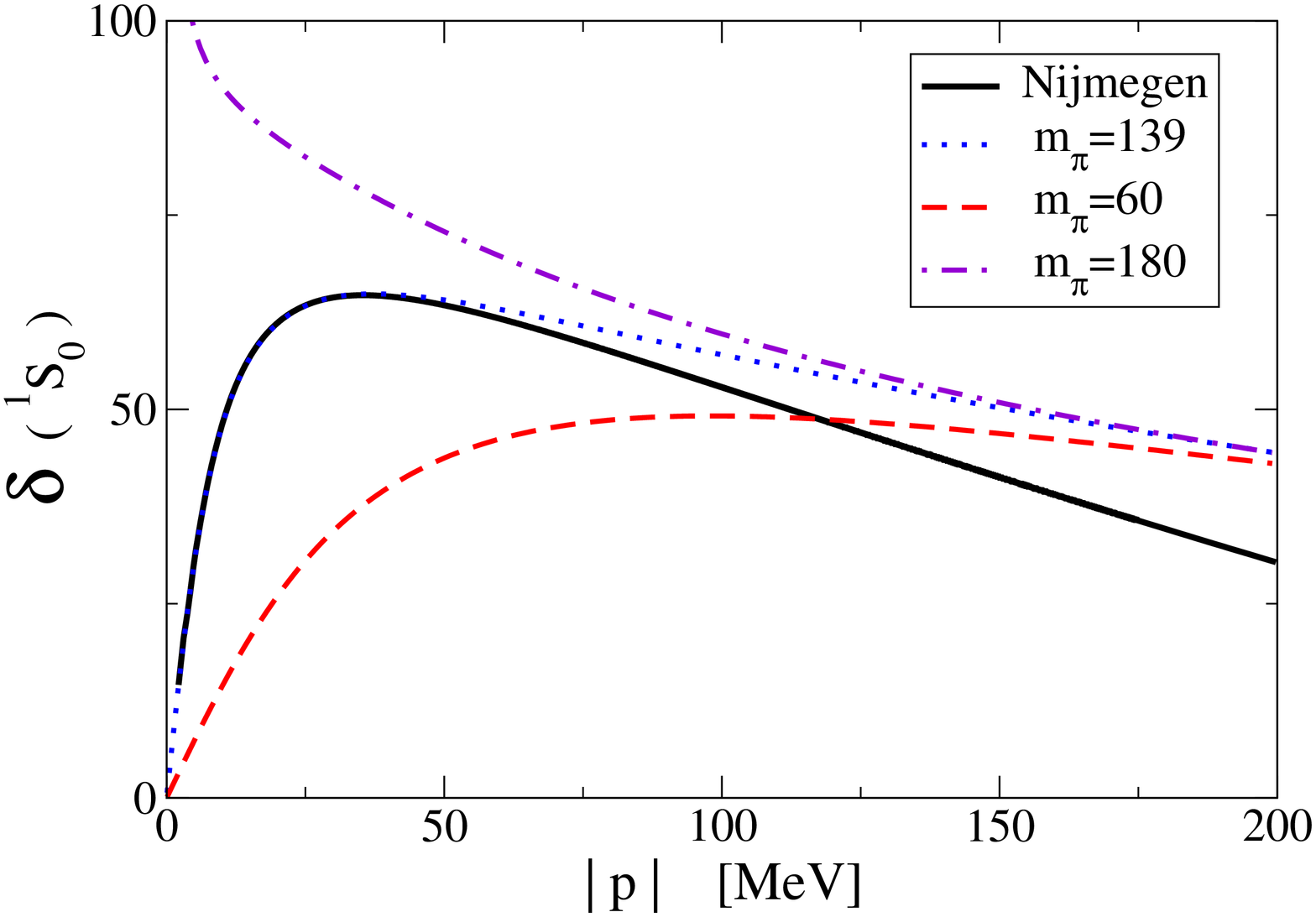,width=3.2in,angle=-90}}
\vspace{0.3cm}
\centerline{\parbox{14cm}{
\caption[fig1]{\label{fig:singrandom}
Properties of the $\si$ channel for the couplings 
in eq.~(\protect\ref{eq:singranpars}).
The left panel shows the scattering length as a function of the pion mass.
For these couplings, there is a window 
where the di-neutron is stable.
The right panel shows the phase-shifts for $m_\pi=60~{\rm MeV}$, 
$m_\pi^{\rm  PHYS}$ and $180~{\rm MeV}$.
}}}
\vspace{0.2cm}
\end{figure}

It is interesting to ask what one might expect if nature has chosen values for
the couplings other than those in eq.~(\ref{eq:singpars}).
Choosing some arbitrary values for the couplings, but ones that still respect
NDA in the two-nucleon sector~\cite{Be01},

\begin{eqnarray}
C_0 (m_\pi^{\rm PHYS}) & = & -4.00~{\rm fm}^2
\quad , \quad 
D_2 (m_\pi^{\rm PHYS}) \ = \ 0.31~{\rm fm}^4
\quad , \quad 
C_2 (m_\pi^{\rm PHYS}) \ =\  4.57~{\rm fm}^4
\ \ \ ,
\label{eq:singranpars}
\end{eqnarray}
we show the scattering length as a
function of the pion mass in Fig.~\ref{fig:singrandom}.
One can see that there are couplings for which the scattering length becomes
positive, indicating the presence of a di-neutron that is stable with respect to the
strong interaction, for relatively small variations in the pion mass.  Perhaps the unjustified choice
of NN interaction arising from N$^2$DA used in previous works on this subject
(e.g. Ref~\cite{dent,fair,shuryak}) is in some way comparable to such an ad-hoc choice of
couplings in the EFT.

It is clear that sufficiently little is known about the 
$m_q$-dependence of the $\si$ channel that placing bounds on the time-variation
of $m_q$ from nuclear processes sensitive to this channel is not possible at this point in time.

\section{The Deuteron Binding Energy}

As we have already discussed, the deuteron plays a central role in BBN and 
the production of light elements.  If the deuteron was deeply
bound or unbound in the nucleosynthesis epoch then the abundances of the light
elements would look radically different from the predictions of BBN and from
those observed in nature.  Therefore, it is possible 
that limits can be placed upon the
time-variation of fundamental couplings if the dependence of 
$B_{\rm d}$, and relevant nuclear cross-sections on these couplings is known.
First realistic attempts to understand $B_{\rm d}(m_q)$
as a function of $m_q$ were made in Refs~\cite{Be01,miller}.
One motivation for BBSvK was to understand just what was possible for the
$m_q$-dependence of $B_{\rm d}(m_q)$, and in particular to determine if the
deuteron was bound or unbound in the chiral limit.
This issue could not be resolved due to the lack of information about 
the $D_2$ operators that contribute to the $\siii-\diii$ coupled-channels.
A second motivation was to understand what was required to make 
any sort of contact with quenched lattice QCD calculations~\cite{fuku}
of the scattering lengths in the $\siii$ channel.

\begin{figure}[!ht]
\centerline{{\epsfxsize=3.63in \epsfbox{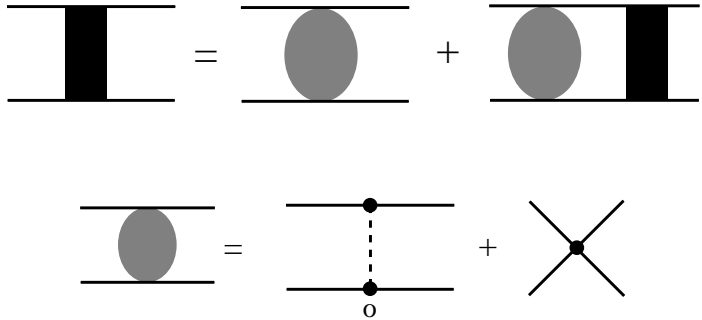}} }
\vskip 0.15in
\noindent
\caption{Lippmann-Schwinger equation for the LO contribution to the
  scattering amplitude (large solid rectangle) in the $\siii -\diii$ coupled-channels. The small solid
  circles denote an insertion of $C_0$ or $g_A$. The ``o'' appearing below the
  OPE diagram implies keeping only the chiral limit.}
\label{fig:3S1LO}
\vskip .2in
\end{figure}

\begin{figure}[!ht]
\centerline{{\epsfxsize=5in \epsfbox{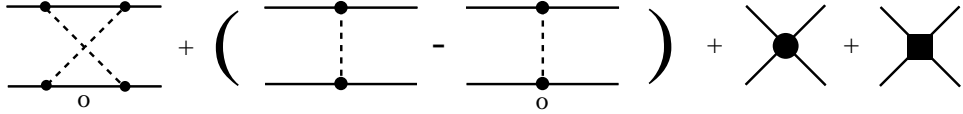}} }
\vskip 0.15in
\noindent
\caption{Chiral limit of the crossed TPE diagram, deviations from the chiral
  limit of OPE, and the $C_2$
  (large solid circle) and $D_2$ (large solid square) operators, all of which
  contribute at NLO in the $\siii -\diii$ coupled-channels. The ``o''
  appearing below a diagram implies keeping only the chiral limit.}
\label{fig:3S1NLO}
\vskip .2in
\end{figure}

In the $\siii-\diii$ coupled-channels, OPE generates both central and tensor
potentials,
\begin{eqnarray}
V_C^{(\pi)} (r;\mpi) & = & -{\alpha_\pi}\,\mpis\;{e^{-\mpi r}\over r} 
\quad , \quad
V_T^{(\pi)} (r;\mpi) \ =\ -{\alpha_\pi}\ {e^{-\mpi r}\over r}
\left(\ {3\over r^2} + {3 m_\pi \over r}\ +\ m_\pi^2\ \right)
\ \ ,
\label{eq:OPEpots}
\end{eqnarray}
where ${\alpha_\pi}= \gas(1-2m_\pi^2 \overline{d}_{18}/\ga)^2/(8\pi f_\pi^2)$. 
The constant $\overline{d}_{18}$ is uncertain~\cite{ulfioffe} but as our
results are quite insensitive to this quantity, we set it to
zero in what follows. At LO in BBSvK counting we should only keep the chiral limit 
of these potentials
and the $C_0$ operator~\footnote{In contrast with the $\si$ channel, in the
$\siii-\diii$ coupled-channels there is a local four-nucleon
interaction from OPE that we have included in our definition of $C_0$.} 
that contributes to the $\siii-\diii$ coupled-channels,
as shown in Fig.~\ref{fig:3S1LO}.
Deviations from the chiral limit are inserted in perturbation theory. 
We choose to keep the full potentials in our  NLO calculation,
meaning that we also include part of the N$^2$LO calculation,
which we find does not modify the divergence structure.
In addition, at NLO in BBSvK counting there is a contribution from the chiral
limit of two-pion exchange (TPE) and from an insertion of $C_2$ and $D_2$, 
as shown in Fig.~\ref{fig:3S1NLO}.
The TPE potential in coordinate space has been computed in Ref.~\cite{KBW}, and
in the chiral limit is given by
\begin{eqnarray}
V_C^{(\pi\pi)} (r; 0) & = & {3(22 g_A^4- 10 g_A^2-1)\over 64\pi^3 f_\pi^4} {1\over r^5}
\quad , \quad
V_T^{(\pi\pi)} (r; 0)  \ =\ -{15 g_A^4\over 64\pi^3 f_\pi^4} {1\over r^5}
\ \ .
\label{eq:TPEpots}
\end{eqnarray}
In order to regulate the potentials at the origin, we use a 
spatial square-well of radius $R$~\cite{Sprung,Be01}, where the 
potential outside the square well is 
\begin{eqnarray}
{\cal V}_L(r;\mpi) & = & \left(
\begin{array}{cc}   
  -M_N V_C(r;\mpi)  & -2\sqrt{2}\; M_N V_T(r;\mpi)\\               
-2\sqrt{2}\; M_N V_T(r;\mpi) & 
-M_N\left( V_C(r;\mpi)-2 V_T(r;\mpi)\right)-{6/{r^2}}
\end{array}
\right)
\ \ ,
\label{eq:potout}
\end{eqnarray}
where 
\begin{eqnarray}
V_C (r; m_\pi) & = & V_C^{(\pi)} (r; m_\pi)\ +\ 
V_C^{(\pi\pi)} (r; 0)
\quad , \quad
V_T (r; m_\pi) \ =\  V_T^{(\pi)} (r; m_\pi)\ +\ 
V_T^{(\pi\pi)} (r; 0)
\ \ .
\label{eq:potsum}
\end{eqnarray}
The potential inside the square well is
\begin{eqnarray}
{\cal V}_S(r;m_\pi ,k^2) & = & 
\left(
\begin{array}{cc}   
-M_N (V_{C_0} + V_{D_2} m_\pi^2 +k^2 V_{C_2})  & 0\\               
0     & -M_N (V_{C_0} + V_{D_2} m_\pi^2 +k^2 V_{C_2})-{6/{r^2}}
\end{array}
\right) 
\label{eq:potin}
\end{eqnarray}
where $V_{C_0}$, $V_{D_2}$ and $V_{C_2}$ are constant potentials
corresponding to the
renormalized local operators with coefficients $C_0$, $D_2$ and $C_2$
in the $\siii-\diii$ coupled-channels,
respectively, and again we have used the LO 
relation between the pion mass and $m_q$.
It is important to recall that there is implicit $m_q$-dependence in
this potential due to $g_A$, $M_N$ and $f_\pi$.  For the chiral limit of the OPE
potential, the NLO $m_q$-dependence of these constants must be retained.
However, for the deviations of the OPE potential from the chiral limit, and
also for the chiral limit of the TPE potential it is sufficient (and in fact
necessary to be able to renormalize the theory) to keep only the chiral limit
of these constants.
Defining $\Psi$ to be
\begin{eqnarray}
\Psi (r) & = & \left(\begin{array}{c}
            u(r)\\
            w(r)
            \end{array}\right)
          \ \ \ , 
\end{eqnarray} 
where $u(r)$ is the S-wave wavefunction and $w(r)$ is the D-wave wavefunction,
the regulated Schr\"odinger equation is 
\begin{eqnarray}
{\Psi'' (r)}
\ +\  \left[\  k^2 \ +\  {\cal V}_L(r;\mpi)\ \theta (r-R) 
\ +\ {\cal V}_S(r;m_\pi ,k^2)\ \theta (R-r)\ \right] \Psi (r) & = & 0.  
\end{eqnarray} 
As discussed in BBSvK, we can make an identification between the coefficients
of local operators, $C_i$ and $D_i$, and the constant
potentials of the square-wells
that enter into eq.~(\ref{eq:potin}), $V_{C_i}$ and $V_{D_i}$,
\begin{eqnarray} 
{C_i}\ \delta^{(3)} (r) & \rightarrow & 
{{3{C_i}\  \theta (R-r)}\over 4\pi R^3}\ \equiv\  {V_{C_i}}\ \theta (R-r)  
\nonumber\\
{D_i}\ \delta^{(3)} (r) & \rightarrow & 
{{3{D_i}\  \theta (R-r)}\over 4\pi R^3}\ \equiv\  {V_{D_i}}\ \theta (R-r)  
\ \ \ .
\label{eq:ctovsinglet}
\end{eqnarray}

At the physical value of the pion mass, $C_2$ and a linear combination of
$C_0$ and $D_2$ are fixed to reproduce the deuteron binding energy and 
the effective range in the $\siii$-channel.
Given that it is only a combination of $C_0$ and $D_2$ that we determine, we
have freedom to vary $D_2$ and compensate this by a change in $C_0$, 
as in the $\si$ channel.
\begin{figure}[htb]
\vspace{-0.2cm}
\centerline{\psrotatefirst
\psfig{file=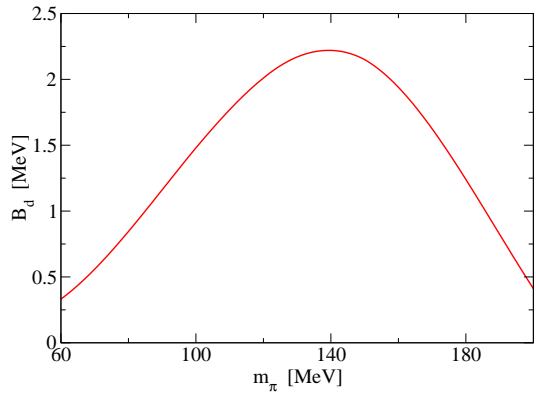,width=3.2in,angle=-90}}
\vspace{0.3cm}
\centerline{\parbox{14cm}{
\caption[fig1]{
\label{fig:deutbind}
The deuteron binding energy at NLO in the EFT as a function of the pion mass for the 
couplings given in eq.~(\protect\ref{eq:trippars}).
For this set of parameters, the deuteron is loosely bound over a wide range of
pion masses.
}}}
\vspace{0.2cm}
\end{figure}
We find couplings $C_0$, $D_2$ and $C_2$, renormalized at the scale set by the
radius of the square-well, $R^*=0.45~{\rm fm}$,
\begin{eqnarray}
C_0 (R^*) & = & -6.17~{\rm fm}^2
\quad , \quad 
D_2 (R^*) \ = \ 0.67~{\rm fm}^4
\quad , \quad 
C_2 (R^*) \ =\  0.75~{\rm fm}^4
\ ,
\label{eq:trippars}
\end{eqnarray}
that are
consistent with NDA and all the low-energy
phase-shift data in the $\siii-\diii$ coupled-channels, and for which the deuteron is loosely bound over a
large range of pion masses, as shown in Fig.~\ref{fig:deutbind}~\footnote{It is
interesting to note that the chiral-limit value of the deuteron binding energy
in the EFT coincides with that found in an analogous calculation with the
Argonne V18 potential, when all coupling constants are frozen to their
physical values, and only the explicit $m_\pi$ dependence in eq.~(\ref{eq:OPEpots})
is considered~\cite{bob}.}.

\begin{figure}[htb]
\vspace{-0.2cm}
\centerline{\psrotatefirst
\psfig{file=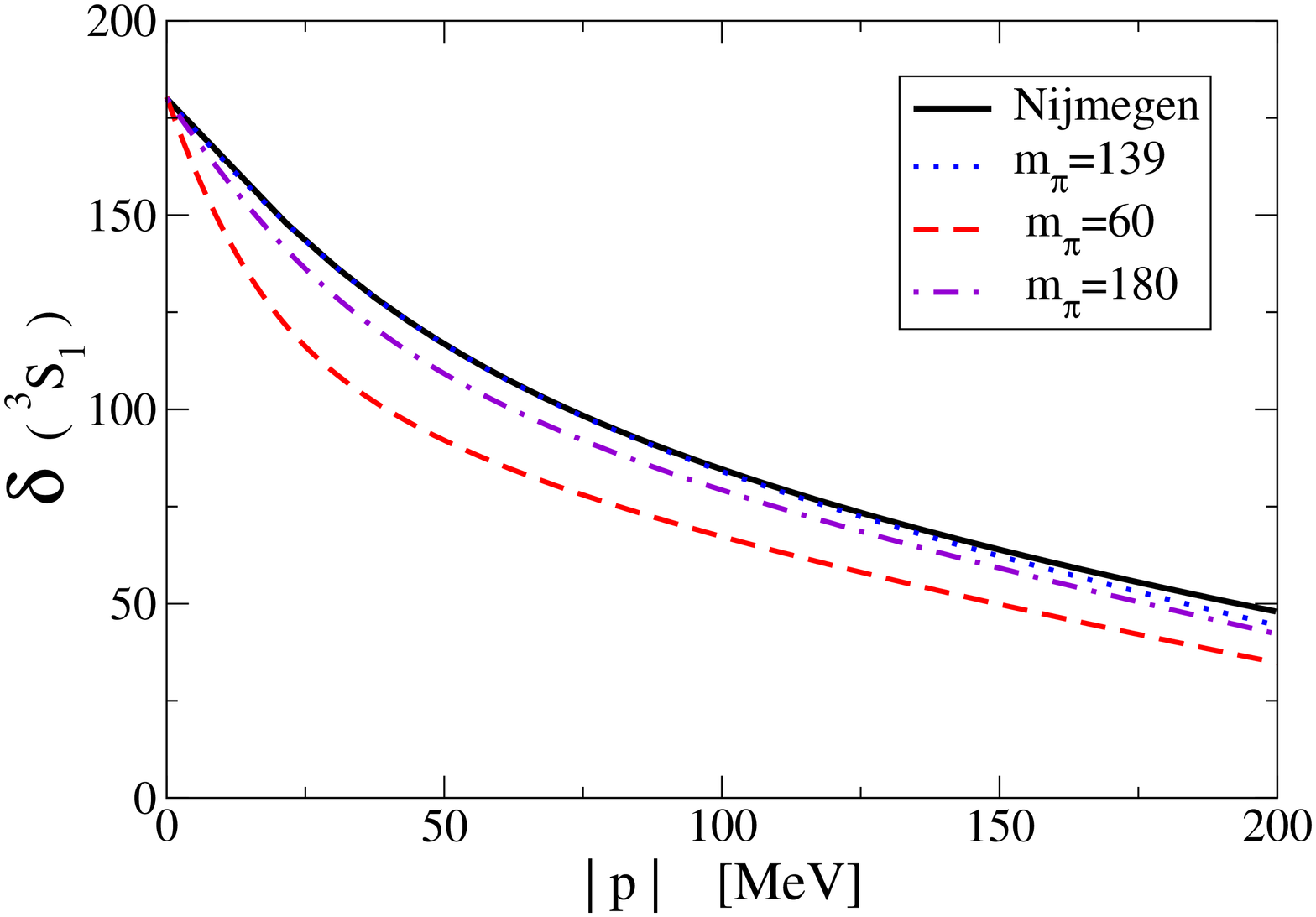,width=2.15in,angle=-90}
\psfig{file=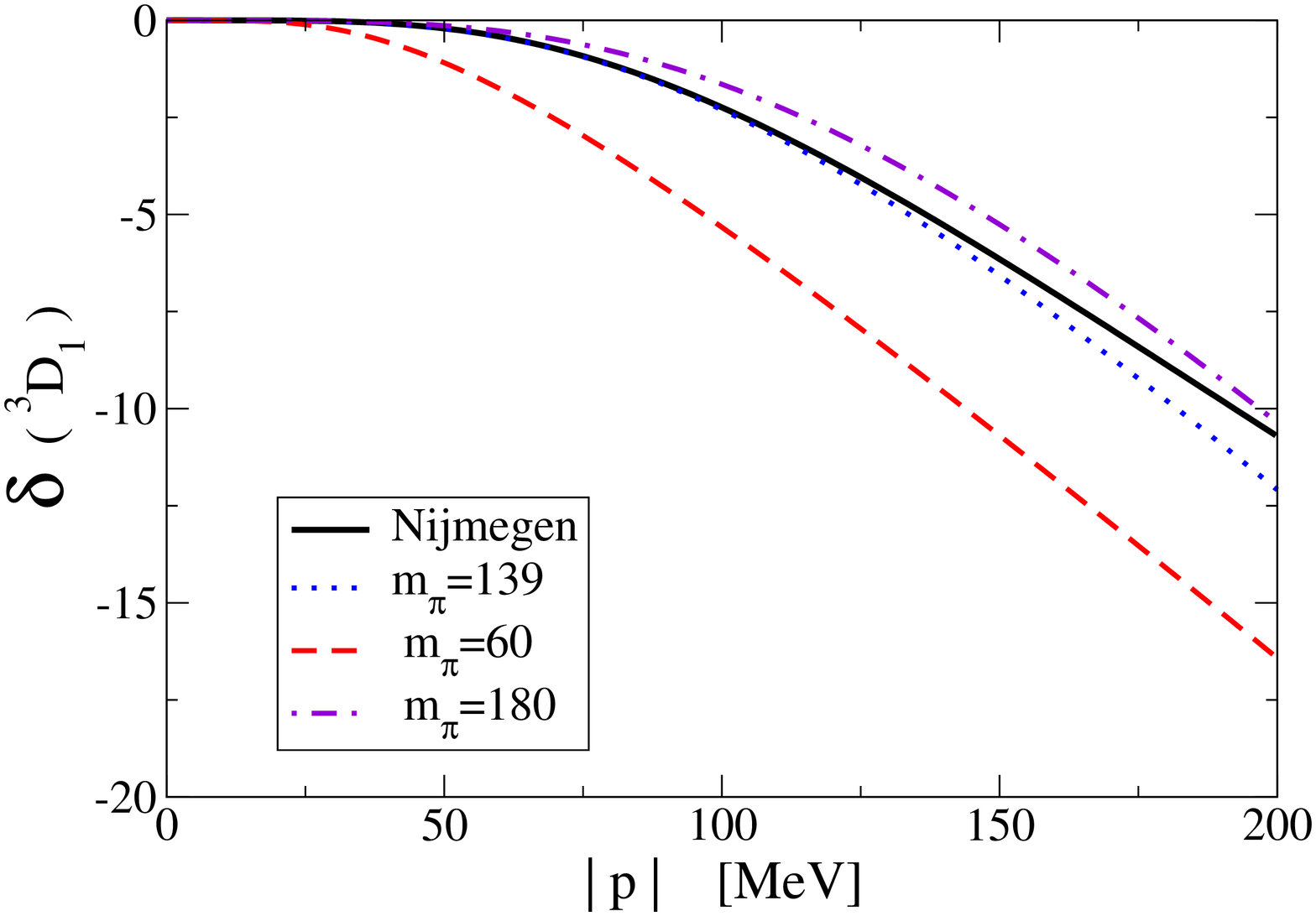,width=2.15in,angle=-90}
\psfig{file=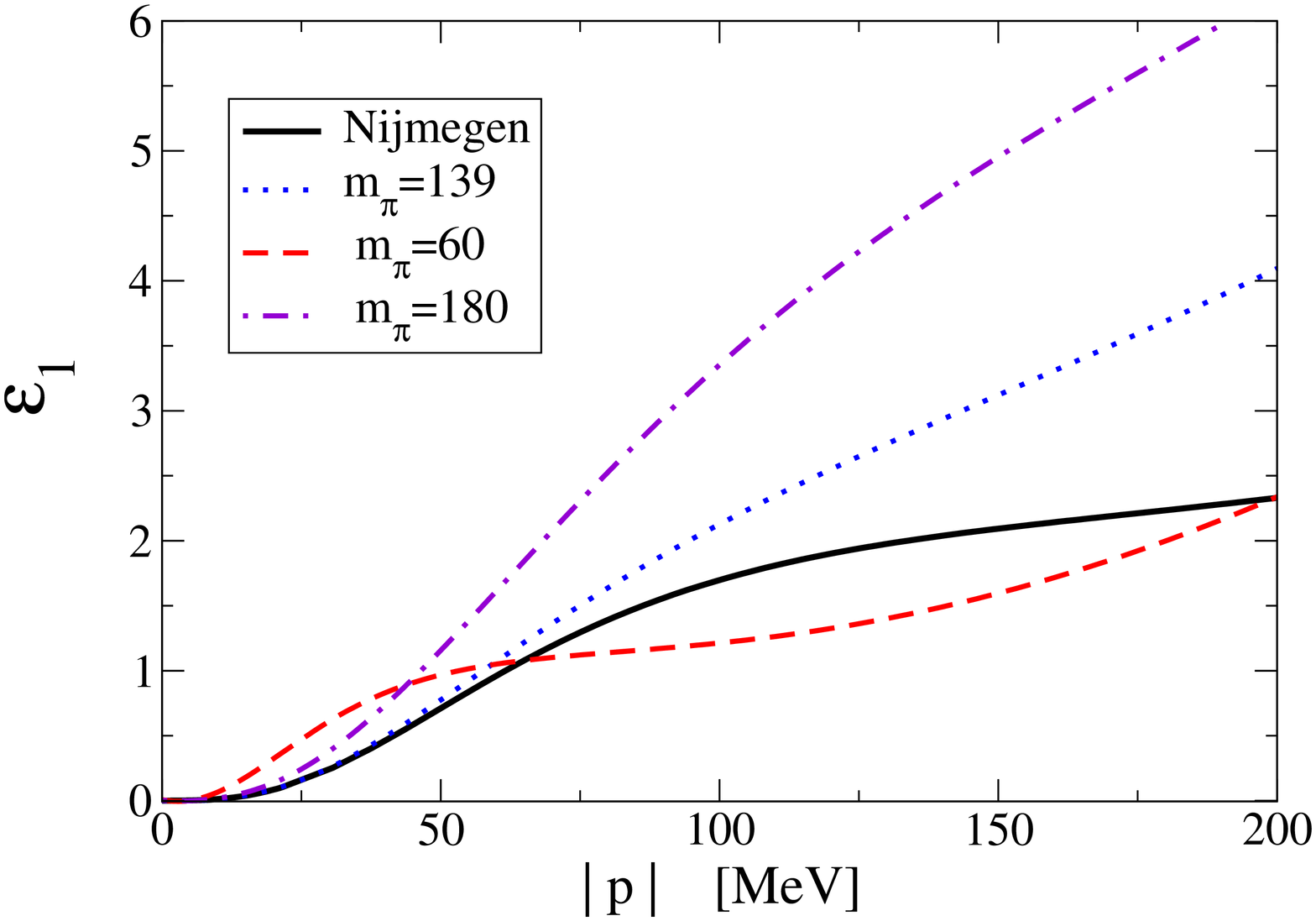,width=2.15in,angle=-90}
}
\vspace{0.3cm}
\centerline{\parbox{14cm}{
\caption[fig1]{\label{fig:tripphase}
The phase-shifts, $\delta_0$ and $\delta_2$ for the $\siii$ channel and the
$\diii$ channel and the mixing parameter $\varepsilon_1$ 
as a function of momentum, $|{\bf p}|$,
for pion masses of $m_\pi=60~{\rm MeV}$ (dashed), 
$m_\pi^{\rm PHYS}$ (dotted) and $180~{\rm MeV}$ (dot-dashed),
for the couplings in eq.~(\protect\ref{eq:trippars}).
The solid curve corresponds to the results of the 
Nijmegen partial-wave analysis~\protect\cite{Nijmegen}.
Note that for $\delta_0$ at $m_\pi^{\rm PHYS}$, the NLO EFT calculation coincides
with the partial-wave analysis to relatively high momenta.
}}}
\vspace{0.2cm}
\end{figure}
The phase-shifts and mixing parameter for the $\siii-\diii$ coupled channels
are shown in Fig.~\ref{fig:tripphase}.
As expected, the $\siii$ phase-shift falls more steeply 
as the deuteron becomes more loosely bound.
Further, as the pion becomes lighter, the higher partial waves and mixing
parameter are expected to increase due to the longer range of the potential
and their relative insensitivity to short-distance physics.

\begin{figure}[htb]
\vspace{-0.2cm}
\centerline{\psrotatefirst
\psfig{file=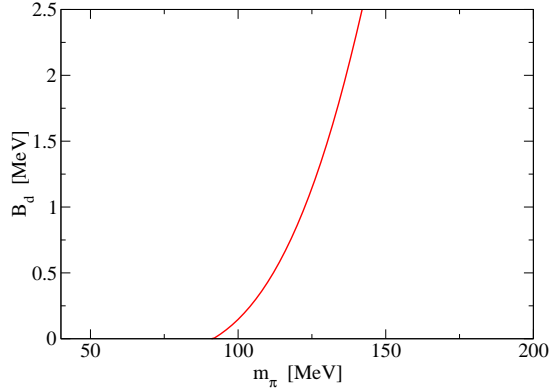,width=3.2in,angle=-90}}
\vspace{0.3cm}
\centerline{\parbox{14cm}{
\caption[fig1]{
\label{fig:deutbindrapid}
The deuteron binding energy at NLO in the EFT as a function of the pion mass for the 
couplings given in eq.~(\protect\ref{eq:triprand}).
}}}
\vspace{0.2cm}
\end{figure}

In BBSvK, the renormalization scale dependence of the theory was
investigated~\cite{Be01}. The dependence on the width of the square-well, $R$,  
was found to be small, consistent with the EFT expectations. We find the same
to be true at NLO.

\begin{figure}[htb]
\vspace{-0.2cm}
\centerline{\psrotatefirst
\psfig{file=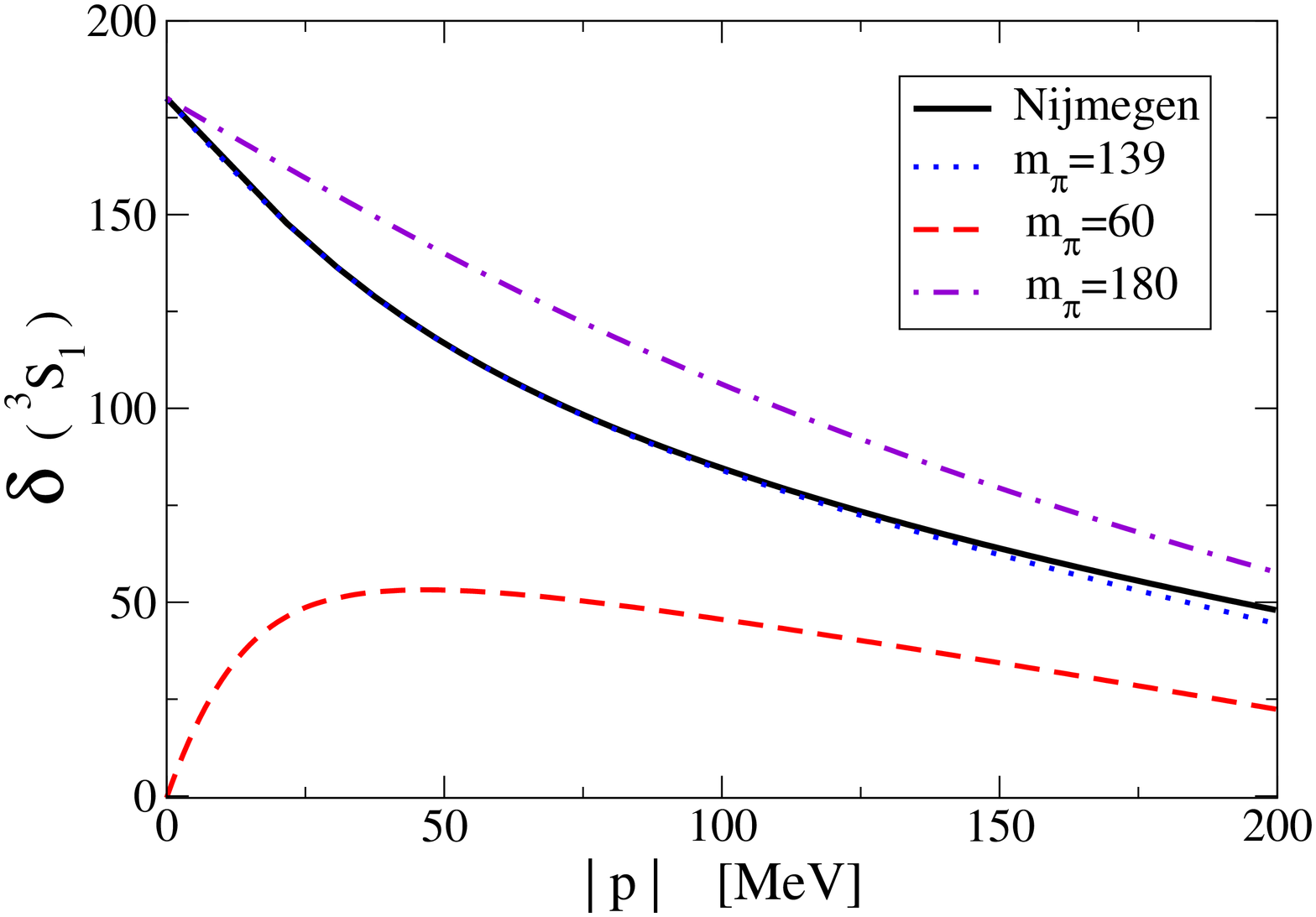,width=2.15in,angle=-90}
\psfig{file=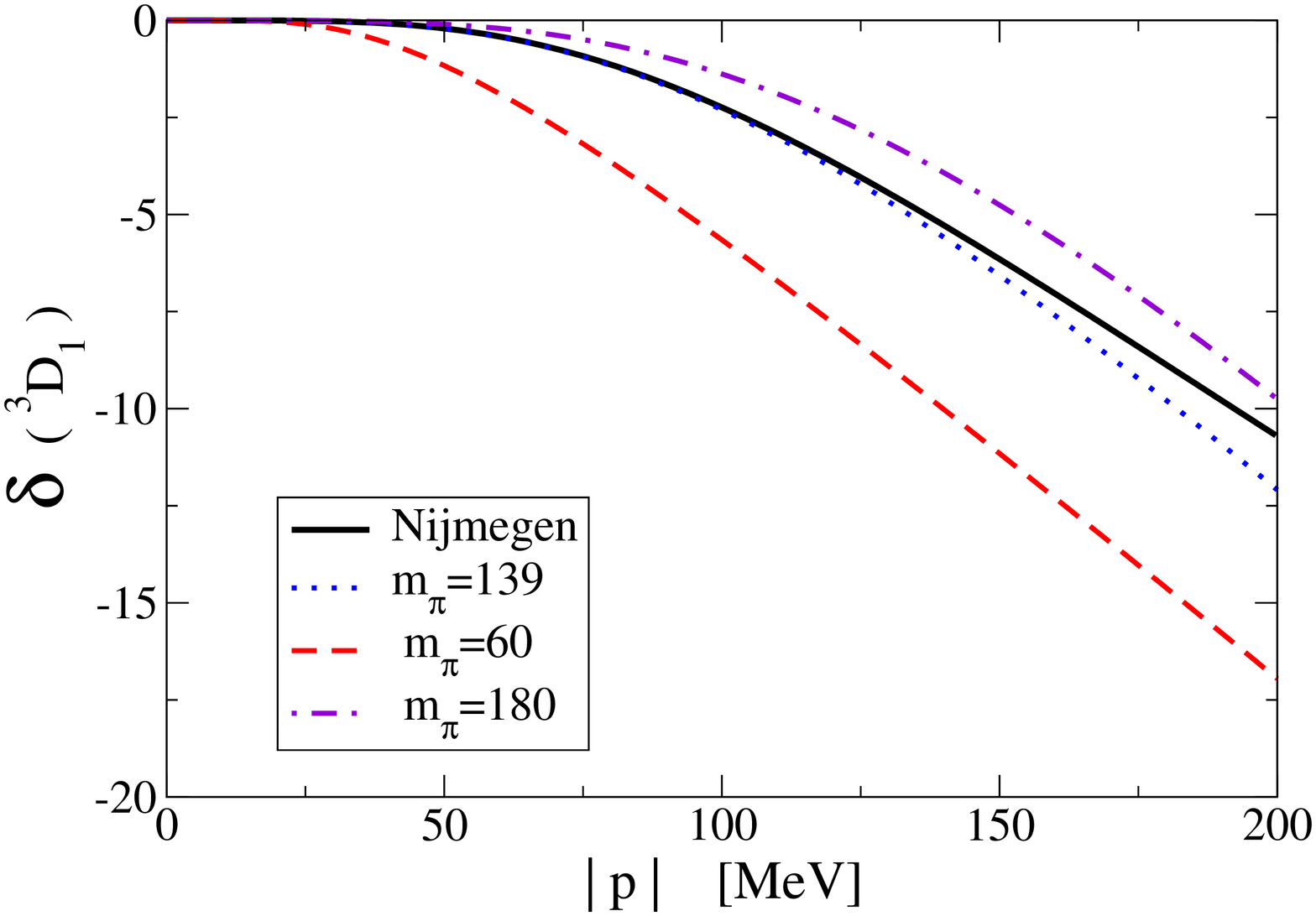,width=2.15in,angle=-90}
\psfig{file=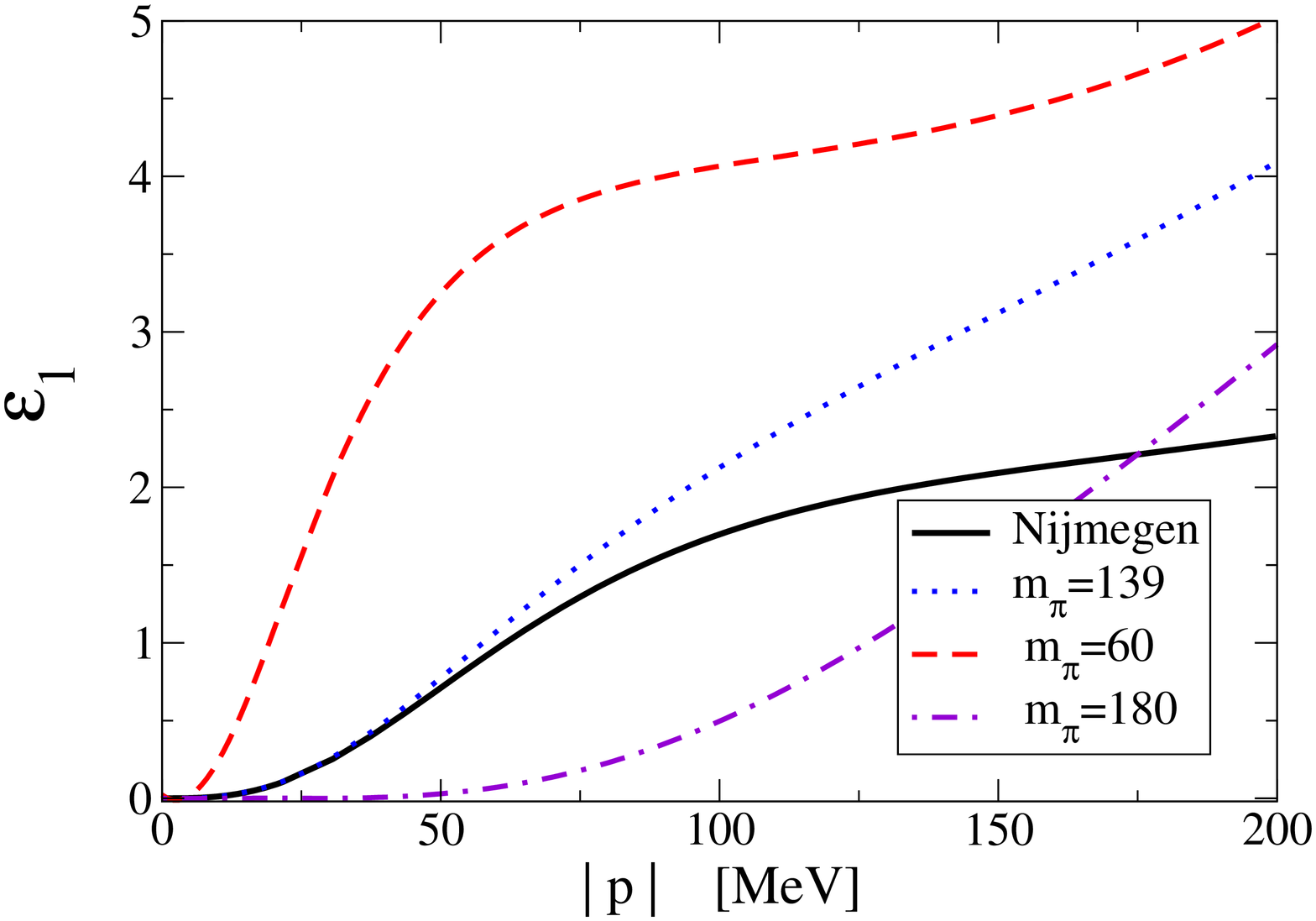,width=2.15in,angle=-90}
}
\vspace{0.3cm}
\centerline{\parbox{14cm}{
\caption[fig1]{\label{fig:tripphaseFAST}
The phase-shifts, $\delta_0$ and $\delta_2$ for the $\siii$ channel and the
$\diii$ channel and the mixing parameter $\varepsilon_1$ 
as a function of momentum, $|{\bf p}|$,
for pion masses of $m_\pi=60~{\rm MeV}$ (dashed), 
$139~{\rm MeV}$ (dotted) and $180~{\rm MeV}$ (dot-dashed),
for the couplings in eq.~(\protect\ref{eq:trippars}).
The solid curve corresponds to the results of the 
Nijmegen partial-wave analysis~\protect\cite{Nijmegen}.
}}}
\vspace{0.2cm}
\end{figure}

For purposes of contrast, we present the deuteron binding energy, phase-shifts and mixing
parameter for an arbitrary set of couplings that respect NDA,
\begin{eqnarray}
C_0 (R^*) & = & -5.05~{\rm fm}^2
\quad , \quad 
D_2 (R^*) \ = \ -1.60~{\rm fm}^4
\quad , \quad 
C_2 (R^*) \ =\  0.75~{\rm fm}^4
\ .
\label{eq:triprand}
\end{eqnarray}
For these choices, the variation of $B_{\rm d}(m_q)$ with respect to 
$m_q$ is quite rapid, as shown in Fig.~\ref{fig:deutbindrapid},
and for $m_\pi \lsim 90~{\rm MeV}$, the deuteron is unbound.
The phase-shifts and mixing parameter resulting from the couplings in 
eq.~(\ref{eq:triprand}) are shown in Fig.~\ref{fig:tripphaseFAST},
and as expected, they have a relatively strong dependence upon $m_q$. 

It is important to point out that while the deuteron binding energy curves
shown in Fig.~\ref{fig:deutbind} and Fig.~\ref{fig:deutbindrapid} 
yield unbound deuterons in the chiral limit, there
exist parameter sets consistent with NDA for which the deuteron remains
bound in the chiral limit.

\section{Conclusion}

We have explored the light-quark mass dependence of low-energy nucleon-nucleon
interactions.  The motivation for this work was to determine if, in fact,
bounds could be set on the time-variation of fundamental couplings from nuclear
processes, such as those occurring during big bang nucleosynthesis.  We have
demonstrated the existence of sets of strong interaction couplings in the
low-energy effective field theory describing the nucleon-nucleon interaction
that are consistent with all available data and with naive
dimensional analysis for which the di-neutron remains unbound, and the deuteron
remains loosely bound over a wide range of light quark masses.  We do not mean
to imply that these are the sets of couplings that nature has chosen, but rather
that this scenario is not excluded at present.  Thus, we conclude that bounds
that have previously been set on the time-variation of fundamental couplings
from processes in the two-nucleon sector are not rigorous and should be discarded.

Our calculation does suffer from some limitations.  First, we have not included
electromagnetism, and thus could not address the $pp$-system in the
$\si$-channel.  However, we do not believe that its inclusion will modify the
results we have presented in any significant way. The same we believe to be
true of isospin breaking. Second, and perhaps the most
important limitation, is that we have not included the strange quark in our
discussions, and have worked with $SU(2)_L\otimes SU(2)_R$ chiral symmetry.
Variations in the strange (and charm, bottom and top) quark mass will manifest 
themselves as changes in the values of the coefficients in the Lagrange density,
$C_i$ and $D_i$.  At this point in time the tools do not exist to analyze this
scenario.  For instance, one might consider developing an EFT with three 
light-quark flavors. However, since nuclei are such finely-tuned systems, and
$SU(3)$-breaking is not so small, it is likely that a perturbative treatment
of nuclei using the approximate $SU(3)_L\otimes SU(3)_R$ chiral
symmetry of QCD would converge very slowly, if at all.

From a conventional nuclear
physics point of view, the $m_q$-dependence of the nuclear potentials
not only requires knowledge of the $m_q$-dependence of one-pion exchange,
but also of the $m_q$-dependence of the
$\rho$, $\omega$, $\phi$,...  masses and their couplings to 
nucleons. This would appear to be an intractable problem.
It is possible that some of the traditional treatments of the
nucleon-nucleon interaction~\cite{Nijmegen,WSS,Pudliner:1997ck,Timmermans}, and treatments of light
nuclei could mimic these dependences by variations in their ad-hoc
short-distance components of the nucleon-nucleon and higher-body potentials.  The recently
developed effective field theory tools organize this problem in a very simple
way, and thus at NLO in the EFT expansion, only the $D_i$ are needed. 
In experimental processes examined to date that, in principle, allow one to separate the $D_i$ from the 
$C_i$, it is found that there are other amplitudes contributing to the process that dominate over the
$D_i$ contributions, rendering such a separation exceedingly difficult.
In the absence of an experimental determination of the $D_i$,
it would appear that the only viable means for determination of this
$m_q$-dependence is through lattice QCD simulations~\footnote{A first attempt 
  at computing nucleon-nucleon scattering
  lengths in quenched lattice QCD has been made in Ref.~\cite{qlattice}.
  A discussion of the two-nucleon potential in quenched and partially-quenched 
  QCD can be found in Ref.~\cite{BSpot}. For a recent discussion, see Ref.~\cite{richards}.}.   
We find this to be very strong motivation to pursue a lattice QCD program focused on the
two-nucleon sector.

\vskip0.2in
\noindent {\large\bf Acknowledgements}

\noindent
We thank George Fuller for useful discussions and Iain Stewart for helpful
comments on the manuscript.
We are especially grateful to Evgeni Epelbaum for pointing out an error in
an earlier version of the manuscript, and for useful discussions.
This research was supported in part by the DOE grant DE-FG03-97ER41014.


\vspace{1cm}
                  
\begin{references}


\bibitem{dalpha}
J.K.~Webb {\it et al},
{\it Phys. Rev. Lett.} {\bf 87}, 091301 (2001);
{\it Mon. Not. Roy. Astron. Soc.}  {\bf 327}, 1208 (2001).

\bibitem{CF}
X.~Calmet and H.~Fritzsh,
{\tt hep-ph/0112110}.

\bibitem{matt}
P.~Langacker, G.~Segre, M.J.~Strassler,
{\it Phys. Lett.} {\bf B528}, 121 (2002).

\bibitem{Chacko:2002mf}
Z.~Chacko, C.~Grojean and M.~Perelstein,
{\tt hep-ph/0204142}.

\bibitem{KPW}
E.W.~Kolb, M.J.~Perry, T.P.~Walker,
{\it Phys. Rev.} {\bf D33}, 869 (1986).

\bibitem{Barrow}
J.D.~Barrow,
{\it Phys. Rev.} {\bf D35}, 1805 (1987).

\bibitem{Dixit}
V.V.~Dixit and M.~Sher,
{\it Phys. Rev.} {\bf D37}, 1097 (1988).

\bibitem{campbell}
B.A.~Campbell and K.A.~Olive,
{\it Phys. Lett.} {\bf B345}, 429 (1995).

\bibitem{Damour}
T.~Damour and F.~Dyson,
{\it Nucl. Phys.} {\bf B480}, 37 (1996).

\bibitem{agrawal}
V.~Agrawal, S.M.~Barr, J.F.~Donoghue, D.~Seckel,
{\it Phys.Rev.} {\bf D57}, 5480 (1998).

\bibitem{hummer}
H.~Oberhummer, R.~Pichler and A.~Csoto,
{\tt nucl-th/9810057}.

\bibitem{Fujii}
Y.~Fujii {\it et al},
{\it Nucl. Phys.} {\bf B573}, 377 (2000);
{\tt hep-ph/0205078}.

\bibitem{Shera}
T.E.~Jeltema and M.~Sher,
{\it Phys. Rev.} {\bf D61}, 017301 (1999).

\bibitem{BIR}
L.~Bergstrom, S.~Iguri and H.~Rubinstein,
{\it Phys. Rev.} {\bf D60}, 045005 (1999).

\bibitem{Csoto}
A.~Csoto, H.~Oberhummer and H.~Schlattl,
{\it Nucl. Phys.} {\bf A688}. 560 (2001);
{\it Nucl. Phys.} {\bf A689}, 269 (2001). 

\bibitem{Chiba}
T.~Chiba,
{\tt gr-qc/0110118}.

\bibitem{dent}
T.~Dent and M.~Fairbairn,
{\tt hep-ph/0112279}.

\bibitem{fair}
M.~Fairbairn,
{\tt hep-ph/0205078}.

\bibitem{Dolgov}
A.D.~Dolgov,
{\tt hep-ph/0201107}.

\bibitem{IK}
K.~Ichikawa and M.~Kawasaki,
{\tt hep-ph/0203006}.

\bibitem{shuryak}
V.V.~Flambaum and E.V.~Shuryak,
{\it Phys. Rev.} {\bf D65}, 0103503 (2002).

\bibitem{olive}
K.A.~Olive, M.~Pospelov, Y.-Z.~Qian, A.~Coc, M.~Casse, 
and E.~Vangioni-Flam, {\tt  hep-ph/0205269}.

\bibitem{uzan}
J.P.~Uzan,
{\tt hep-ph/0205340}.

\bibitem{We90}
S.~Weinberg,
{\it Phys. Lett.} {\bf B251}, 288 (1990);
{\it Nucl. Phys.} {\bf B363}, 3 (1991);
{\it Phys. Lett.} {\bf B295}, 114 (1992).

\bibitem{KSWa} 
 D.B.~Kaplan, M.J.~Savage, and M.B.~Wise, 
{\it Nucl. Phys.} {\bf B478} 629 (1996).

\bibitem{KSWb}
 D.B.~Kaplan, M.J.~Savage, and M.B.~Wise, 
{\it Phys. Lett.} {\bf B424}, 390 (1998); 
{\it Nucl. Phys.} {\bf B534}, 329 (1998).


\bibitem{Be01}
S.R.~Beane, P.F.~Bedaque, M.J.~Savage and U.~van Kolck,
{\it Nucl. Phys.} {\bf A700}, 377 (2002).

\bibitem{Birse}
M.C.~Birse, J.A.~McGovern and K.G.~Richardson,
{\it Phys. Lett.} {\bf B464}, 169 (1999).

\bibitem{FMS} 
S.~Fleming T.~Mehen and I.W.~Stewart, 
{\it Nucl. Phys.} {\bf A677}, 313 (2000).

\bibitem{Be00}
S.R.~Beane, P.F.~Bedaque, W.C.~Haxton, D.R.~Phillips and 
M.J.~Savage, Essay for the Festschrift in honor of Boris Ioffe, in the
Encyclopedia of Analytic QCD, 
At the Frontier of Particle Physics, vol.~1,  133-269, 
edited by M.~Shifman (World Scientific); {\tt nucl-th/0008064}.

\bibitem{pablito}
P.F.~Bedaque and U.~van Kolck,
{\tt nucl-th/0203055}.

\bibitem{danielito}
D.R.~Phillips,
{\tt nucl-th/0203040}. 

\bibitem{ulfito}
E.~Epelbaum, U.-G.~Mei\ss ner, W.~Gl\"ockle, C.~Elster, H.~Kamada, A.~Nogga and H.~Witala,
{\tt nucl-th/0109065}.

\bibitem{parkito}
T.S.~Park, K.~Kubodera, D.P.~Min and M.~Rho,
{\it Nucl. Phys.} {\bf A684}, 101 (2001).

\bibitem{kiddies}
S.R.~Beane, P.F.~Bedaque, L.~Childress, A.~Kryjevski, J.~McGuire 
and U.~van~Kolck,
{\it Phys. Rev.} {\bf A64}, 042103 (2001).

\bibitem{ulfioffe}
U.-G.~Mei\ss ner,
Essay for the Festschrift in honor of Boris Ioffe, 
in the Encyclopedia of Analytic QCD, 
At the Frontier of Particle Physics, vol.~1, 417-505, edited by M.~Shifman
(World Scientific); {\tt hep-ph/0007092}.

\bibitem{GaLe84}
J.~Gasser and H.~Leutwyler,
{\it Annals Phys.} {\bf 158}, 142 (1984). 

\bibitem{CoGaLe01}
G.~Colangelo, J.~Gasser and H.~Leutwyler,
{\it Nucl. Phys.} {\bf B603}, 125 (2001)

\bibitem{FeMe00}
N.~Fettes and U.-G.~Mei\ss ner,
{\it Nucl. Phys.} {\bf A676}, 311 (2000). 

\bibitem{carlos}
C.~Ordo\~nez, L.~Ray and U.~van Kolck,
{\it Phys. Rev.} {\bf C53}, 2086 (1996);
{\it Phys. Rev. Lett.}  {\bf 72}, 1982 (1994);
{\it Phys. Lett.} {\bf B291}, 459 (1992).

\bibitem{Nijmegen}
V.G.J.~Stoks, R.A.M.~Klomp, M.C.M.~Rentmeester, and J.J.~de~Swart,
{\it Phys. Rev.} {\bf C48}, 792 (1993).

\bibitem{cohen}
T.D.~Cohen and J.M.~Hansen,
{\it Phys. Rev.} {\bf C59}, 13 (1999);
{\it Phys. Rev.} {\bf C59}, 3047 (1999).

\bibitem{miller}
A.~Bulgac, G.A.~Miller and M.~Strikman,
{\it Phys. Rev.} {\bf C56}, 3307 (1997).

\bibitem{fuku}
M.~Fukugita, Y.~Kuramashi, M.~Okawa, H.~Mino and A.~Ukawa,
{\it Phys. Rev.} {\bf D52}, 3003 (1995).

\bibitem{KBW}
N. Kaiser, R. Brockmann and W. Weise,
{\it Nucl. Phys.} {\bf A625}, 758 (1997).

\bibitem{Sprung}
D.W.L.~Sprung, W.~van Dijk, E.~Wang, D.C.~Zheng,
P.~Sarriguren, and J.~Martorell,
{\it Phys. Rev.} {\bf C49}, 2942 (1994).

\bibitem{bob}
R.B.~Wiringa, {\it private communication}.

\bibitem{WSS}
R.B.~Wiringa, V.G.~Stoks and R.~Schiavilla,
{\it Phys. Rev.} {\bf C51}, 38 (1995).

\bibitem{Pudliner:1997ck}
B.S.~Pudliner, V.R.~Pandharipande, J.~Carlson, 
S.C.~Pieper and R.B.~Wiringa,
{\it Phys. Rev.} {\bf  C56}, 1720 (1997).

\bibitem{Timmermans}
R.G.~Timmermans,
{\it Nucl. Phys.} {\bf A689}, 23 (2001).

\bibitem{qlattice}
M.~Fukugita, Y.~Kuramashi, M.~Okawa, H.~Mino, and A.~Ukawa,
{\it Phys. Rev.} {\bf D52}, 3003 (1995).

\bibitem{BSpot}
S.R.~Beane and M.J.~Savage,
{\it Phys. Lett.} {\bf B535}, 177 (2002).

\bibitem{richards}
D.G.~Richards,
{\tt nucl-th/0011012}.



\end{references}
\end{document}